# MFA is a Waste of Time! Understanding Negative Connotation Towards MFA Applications via User Generated Content

S. Das, B. Wang, and L. Jean Camp
School of Informatics, Computing, and Engineering
Indiana University Bloomington
{sancdas, bw10, ljcamp} @iu.edu

## Abstract

Traditional single-factor authentication possesses several critical security vulnerabilities due to single-point failure feature. Multi-factor authentication (MFA), intends to enhance security by providing additional verification steps. However, in practical deployment, users often experience dissatisfaction while using MFA, which leads to non-adoption. In order to understand the current design and usability issues with MFA, we analyze aggregated user generated comments ($N = 12{,}500$) about application-based MFA tools from major distributors, such as, Amazon, Google Play, Apple App Store, and others. While some users acknowledge the security benefits of MFA, majority of them still faced problems with initial configuration, system design understanding, limited device compatibility, and risk trade-offs leading to non-adoption of MFA. Based on these results, we provide actionable recommendations in technological design, initial training, and risk communication to improve the adoption and user experience of MFA.

## Keywords

Authentication, Multi-Factor Authentication, Passwords, User Studies, User Experience, User Evaluation.

## 1. Introduction

In recent years, exponential growth for both internet users as well as security attacks impacting these users, has been noted. A primary cause for these recent security incidents is contributed from improperly designed or implemented authentication systems. Traditional single-factor authentication, such as passwords, have dominated authentication system design for a long time (Hwang & Li 2000). However, under the increasing complexity of security threats in the internet (Das, Kim, Tingle & Nippert-Eng, 2019), the password model is susceptible to several security vulnerabilities (Joyce 2016). Thus, we cannot rely on a single-factor authentication system for mission-critical sectors such as finance, health care, government, and others (Ward 2006).

As risk mitigation strategies, security researchers often recommend increasing password complexity and using password managers (Choong, Theofanos & Liu 2014; Camp, Abbott & Chen 2016). As an improved solution, Multi-factor authentication (MFA) has been proposed to address the vulnerability of a single-factor authentication system (Amin, ul Haq & Nazir 2017, , Das, Wang, Tingle & Camp, 2019) by adding multiple layers in addition to passwords, such as fingerprints, Face ID, Hardware tokens, etc. to reduce the attack surface (Althobaiti & Mayhew 2014; Ting, Hussain & LaRoche 2015). Despite its benefits, usability of MFA tools remains a challenge which hinders user adoption (Das, Dingman & Camp 2018). In real-world deployments, it is common to observe difficulties in setting up MFA or complainants from a significant portion of users (Furnell 2007). To understand user's perspective about MFA, user generated contents such as reviews can be treated as an important indicator to analyze usability issues (Braz & Robert 2006). Thus, we analyzed user comments of popular MFA application-based software in both consumer and enterprise markets. We found that, irrespective of security concerns, users often have negative attitudes toward MFA adoption.

In section 2, we summarize existing research discussing MFA technologies and effectiveness of user generated comments. Section 3 provides a detailed encounter of the study protocol, followed by the critical findings of the study in section 4. We conclude by annotating crucial issues in current MFA implementation in section 4, and recommendations are made in section 5.

## 2. Related Work

While MFA dramatically improves online security, a slow rate of MFA adoption has been observed due to existing human-centered issues in MFA technologies. Security and usability are both essential in the authentication process (Braz & Robert 2006). Current security and privacy tools, such as Tor (McCoy, Bauer, Grunwald, Tabriz & Sicker 2007), Pretty Good Privacy (PGP) (Whitten & Tygar 1999) and MFA (Armington & Ho 2003), have certain negative impacts on the user experience, thus preventing them from being widely and correctly utilized. Similar occurrence is observed on MFA either.

Braz et al. has pointed out that human factors and the graphical user interface (GUI) impact the overall user experience with multi-factor authentication (Braz & Robert 2006). Das et al. studied the user experience of FIDO U2F's and revealed that issues with enrollment and verification have caused trouble for users choosing the FIDO U2F (Das, Dingman & Camp 2018). Weir et al. conducted experiments with phone-based banking and suggested that additional verification slowed down the banking process (Weir, Douglas, Richardson & Jack 2010). Colnago et al. studied user experience with MFA in the context of organization-wide deployment, such as within universities (Colnago, Devlin, Oates, Swoopes, Bauer, Cranor & Christin 2018). In our research, we present the analysis of multi-factor authentication usability through user reviews from distributors to understand users' ideas for it. In this regard, we first present existing research on the technology itself and its improvements, followed by the instrument used for the comment review.

Users reviews are an important part in mobile application development (Md Noman, Das & Patil 2019). There are existing researches on analyzing user reviews from application distributor to understand users' desire and mental activities for improving product quality and making new feature decisions. Fu et al. studied the case of users that dislike applications and their primary expectation of applications, and proposed the key metrics that users focused on for both mobile applications and games [1] using statistical models (Fu, Lin, Li, Faloutsos, Hong & Sadeh 2013) such as topic models. They identified users' primary considering for choosing applications are price, features and stability. In our work, we used Natural Language Processing (NLP) technologies trained to extract topic words from user reviews and grouped them to understand key metrics that users care about in multi-factor applications.

Topic models and clusters of words were intensively used in analysis text content such as user reviews and status messages. Ding et al. proposed a lexicon-based approach to opinion mining (Ding, Liu & Yu 2008). They performed lexicon-based analysis and transformed review text into score factors to understand topics and features presented in texts. Pang et al. also combined sentiment analysis in such analytics (Pang, Lee et al. 2008). Much work are also focused on removing irrelevant or spam reviews from the whole review data (Jindal & Liu 2008; Li, Huang, Yang & Zhu 2011; Mukherjee, Liu & Glance 2012). In our work, while using widely deployed spam detection system, we use pre-trained text model to perform sentiment analysis as well as keyword extraction for comment data, and then performed text clustering to understand critical issues or needs in multi-factor authentication.

## 3. Methods

The aim of our study was to understand user perception and adoption of MFA. Thus, we targeted anonymous crowd-sourced comments, which included user reviews (Md Noman et al. 2019) of various MFA tools and technologies. MFA methods vary considerably, ranging from app-based MFA tools to hardware tokens to software installed in a trusted device, and others. Many organizations are focusing on building MFA tools, such as Google [2], Microsoft [3], Yubico [4], Okta [5], Duo Security, etc.

### 3.1. Data Collection

Research has shown that the majority of user reviews and comments are typically short and without much information (Vasa, Hoon, Mouzakis & Noguchi 2012). Thus, it was necessary for us to acquire a large amount of data to support content and cluster analysis. Thus, we collected crowd-sourced comments ($N$ = 12500) of five app-based MFA solutions, *Duo Security Phone App Authentication, Google Authenticator, Microsoft Authenticator, and Authy* [6] from the Apple App Store, Google Play Store, Amazon Marketplace, along with internal reviews from organizations that mandate MFA. As mentioned above, we also explored users' comments in scenarios such as organization deployments; hence, we selected an organization that adopted MFA and retrieved data from their internal site for application navigation. To ensure the content quality, we ran additional filters ($N$ = 12500) within the collected dataset using automation tools based on the following criteria:

1. Comments should have at least one complete sentence. Comments that only contained emojis or short word groups (less than 100 characters) were discarded.
2. We also ran these comments through Akismet [7]; an industry-standard service provider for anti-spam solutions, for spam and bot filtering.

---

[1] While games are considered as mobile applications, it is typical to separate them from normal mobile applications in such analytics.

[2] https://www.google.com/landing/2step/

[3] http://aka.ms/azuread

[4] https://yubico.com

[5] https://www.okta.com/

[6] https://authy.com/

The majority of the above-mentioned tools follow the agile method of software development (Beck, Beedle, Van Bennekum, Cockburn, Cunningham, Fowler, Grenning, Highsmith, Hunt, Jeffries et al. 2001), so including various updated and upgraded versions of the tools were important while discussing the user reviews. We wanted to perform version control analysis of the tools in order to understand user feedback through product iterations. As a result, we kept additional meta data (versions, upgrades made, etc.) of the software as well. We will discuss the analysis techniques in the next subsection.

## 3.2. Analysis Technique

We evaluated the above-mentioned MFA tools based on the user ratings and their review comments. Automated text analytic application services, such as the Microsoft Azure Text Analytics API [8] were used to perform analysis of the collected user generated comments. Throughout our analysis, we provided detailed information, such as text language, user emotion, sentiment score, and key content for the user comments. This information, along with other meta data, was utilized in the analysis procedure for extracting critical information from user reviews. We grouped user reviews based on the application version to understand the effects of application or service version iteration, as well as possible improvements or regressions. To better visualize the content, we processed in the analysis procedure, we used data visualization technologies such as word cloud graphs and other data charts to present user opinions of these MFA services. Additionally, we randomly selected user comments ($M$ = 300) from each category to perform qualitative analysis in order to form a deeper understanding of the usability issues with MFA.

# 4. Results and Analysis

We collected user reviews about the different application-based MFA tools through apps stores and marketplace where we also performed keyword clustering. With filtered keywords, we composed a word cloud for proper data visualization of user comments. Figure 1 represents the word cloud of review data, which provides an overview of the comments. Positive terms, such as, 'great', 'best', 'support', etc. can be noted, but without proper context and sentiment analysis might paint out an incorrect picture of MFA usage.

[7] https://akismet.com

[8] https://azure.microsoft.com/en-us/services/cognitive-services/text-analytics/

Thus, we analyzed the sentiment score [0,1] for comments, where 0 represented extremely negative and 1 represented extremely positive. As the Microsoft Azure Cognitive tool 3.2 stated, the comments scoring more than 0.5 to 1 are considered to be positive; otherwise, the comment is considered negative. Combined with the keywords, we identified the following top review topics that represent customers' dissatisfaction, suggestions, and complaints. Noticeably, we found that positive comments are over-generic: a majority of positive comments just rated applications as "Great App" because it was instructed by their organization to use them, however, when delved deeper in the comments, a significantly low population reported the security benefits of such tools. In contrast, negative comments are mostly targeted towards specific issues, such as device incompatibility, lack of user tool understanding, etc. Additionally, we noted that negative sentiments (56.9%) overpowered the positive yet generic connotation towards MFA.

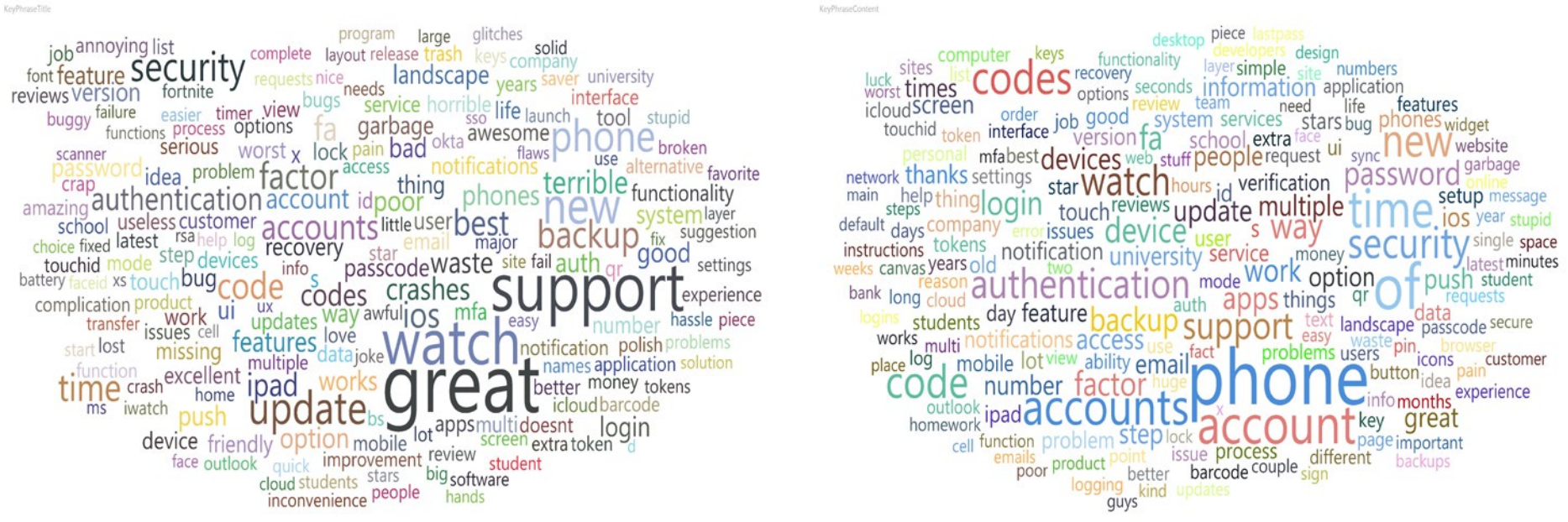

**Figure 1:** Word Cloud for showing the distribution of title (Fig.1) and contents (Fig.2) of the user reviews

We compared overall ratings, as well as grouped user ratings, with the application version to analyze the software development trend. Among all mobile applications, we identified generic one-time password (OTP) applications, such as Authy, to be more popular among users, while service-specific applications were more likely to receive lower scores. While, Authy had the highest average score, and Okta had the lowest score as mentioned in table 1.

| MFA Applications | User Ratings | Sentiment Scores of Review Content |
|---|---|---|
| Authy | 3.866 | 0.579 |
| Microsoft Authenticator | 2.702 | 0.449 |
| Duo Security | 2.356 | 0.349 |
| Google Authenticator | 2.054 | 0.357 |
| Okta | 2.031 | 0.406 |

**Table 1:** Distribution of the MFA applications and their user rating and sentiment scores. Sentiment scores are placed in $[0,1]$, value less than $0.5$ is considered to be negative.

Figure 2 shows that the user reviews for most applications change over a period based on their versions, however, not showing a positive linear acceptability trend. Duo Security and Okta had major review decent during development iterations, Authy had generally higher review scores across iterations, and Microsoft Authenticator and Google Authenticator had minor improvements. Combined with NLP analysis in figures 3 and 4 of sentiment scores, we found a positive correlation between user reviews scores and their comment emotion.

By selecting random comments from each category in the negative comments, we established deeper understanding of issues in current MFA services. We've identified a few major usability issues in these MFA implementations:

1. Backup and Migration: Users expressed concerns about current MFA backup mechanisms. Either no backup is provided, or the backup "routine check" confused users about its security authenticity. Additionally, users are often unable to migrate MFA credentials into new devices, thus showing lack of device backup strategies.

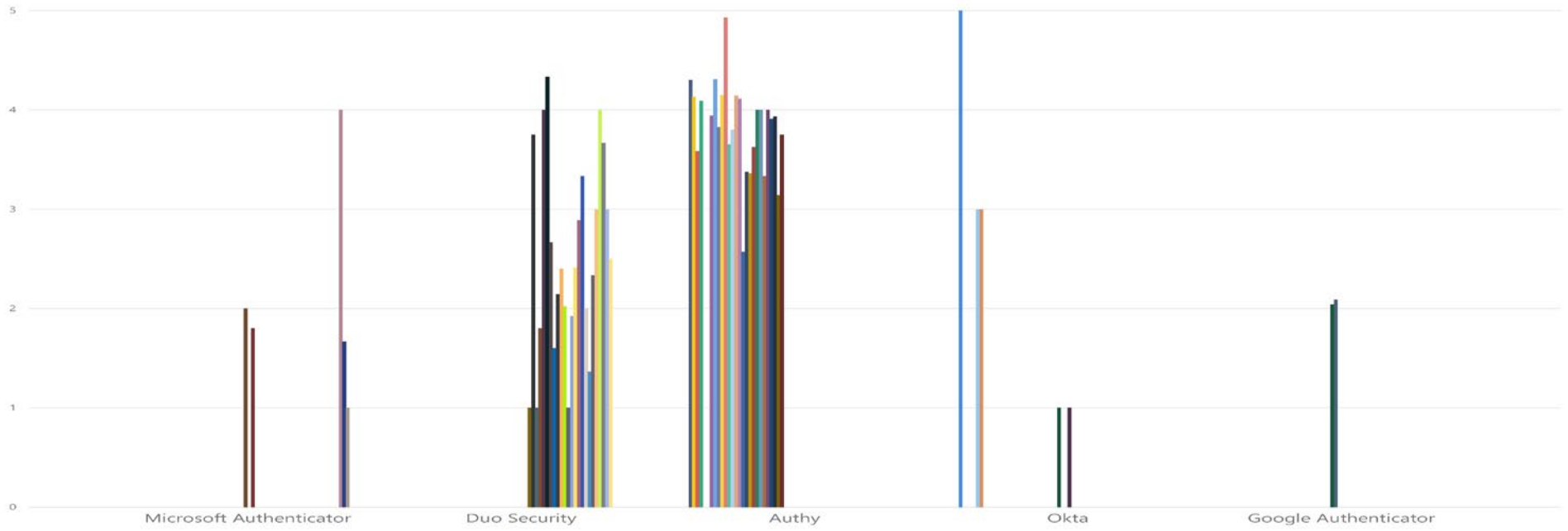

**Figure 2:** Ratings of the different versions of the five applications.

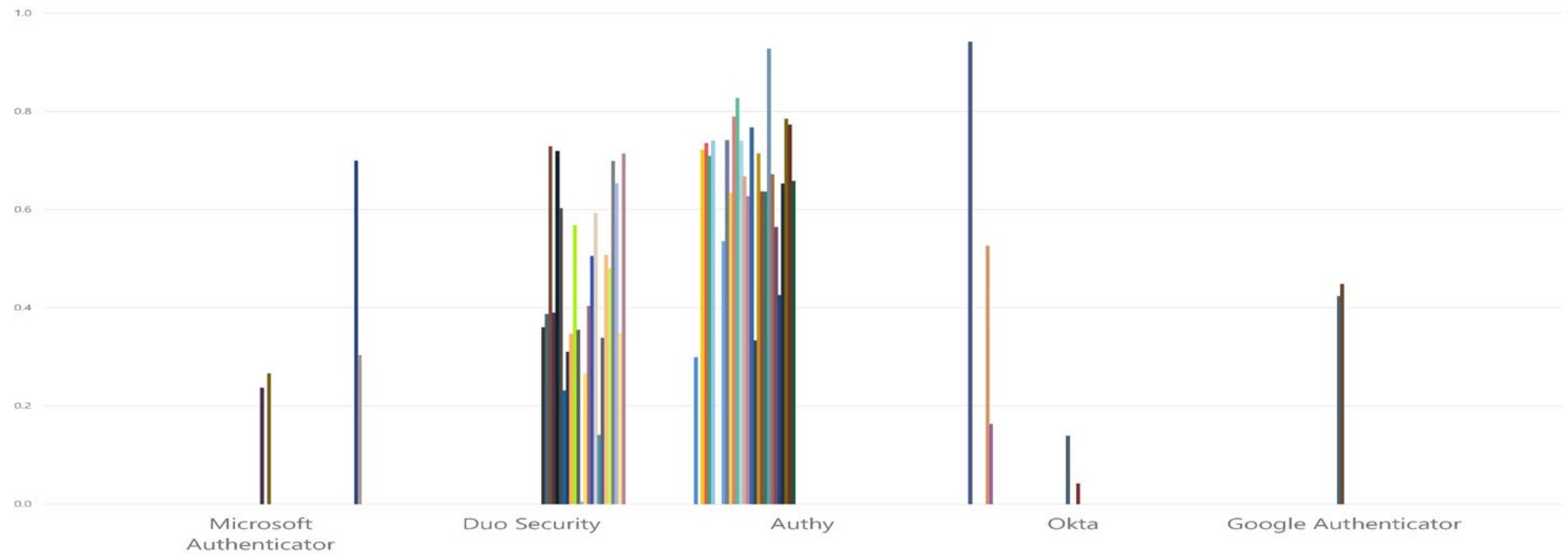

**Figure 3:** Sentiment score for review titles from $[0, 1]$. 0 represents extreme negative emotion while 1 represents the extreme positive motion.

2. Setup, Compatibility, Application and Integration Quality: As technology evolves, users who have adopted to MFA, express higher demands for MFA device compatibility, such as with smart watches. In addition, users expressed difficulties with setting up MFA for the first time. Furthermore, users expressed difficulties with using these applications due to crashes or poorly integrated systems.

3. Forced-to-use: Making technology mandatory often leads to rejection of the same; we see similar trends among MFA users, where they stated not understanding additional security benefits provided by MFA. They complained that they were forced to use MFA by their work or educational organization without explanations.

As for the consideration of disaster recovery, users are always concerned about backup and migration capabilities. Unfortunately, most applications do not properly implement secure backups, or the backup experience occasionally confused users. For example, Google Authenticator intentionally did not store MFA code seeds in cloud by design, since device ownership is treated as a factor. As a result, users have to re-enroll in MFA every time they re-install the application or replace the device:

*...one reason only, that is to preserve my codes ... and it fails. ...to your new one, every single key is broken because it's tied to the device. the keys did not move to my new phone*

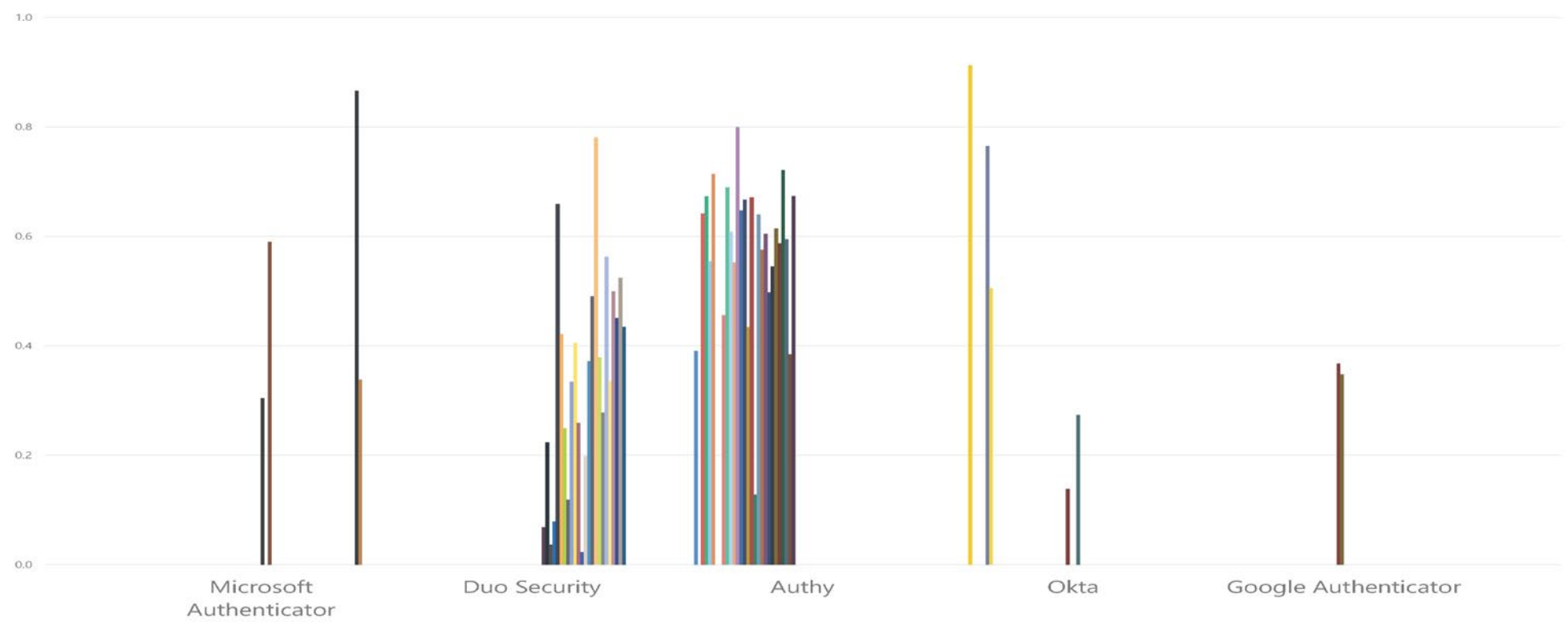

**Figure 4:** Sentiment score for the content of the user reviews valued between $[0, 1]$.

Some applications implement secure backups, but they require users to confirm their backup password routinely, since developers are concerned about users' memorization issues. Unfortunately, this confused and annoyed users:

*I have to verify my backup password as well making this a huge pain. Why extra backup password?*

As technology evolves, users have higher demand for new device support. Sometimes the application support cannot keep up with device updates. As a result, compatibility issues occurred. In addition, poorly integrated online systems harm the overall experience of MFA.

*Please update this app with a watch complication.*
*90% of the time the watch app just says that it couldn't generate a pass code.*
*The problem is that the app now crashes every time I try to open it!*

The lack of proper user training and risk communication caused confusions and misconceptions of MFA. Users complained that they could hardly understand the system design (such as the device factor, as mentioned above) and additional security benefits delivered by MFA. In addition, users expressed hostility with the inconvenience of using MFA, as they were forced to use MFA by their employers and schools.

*Does not work well with Password Managers!*

*the only reason I've downloaded this app is cause X requires it now to login into to everything.*

*My employer uses Duo ...and it is absolutely ... in THE worst security app ever...*

# 5. Implications

Based on our results, we make actionable suggestions for improving user training, risk communication, and application design in order to mitigate usability, adoption, and acceptability issues for better MFA adoption.

## 5.1. User Training and Risk Communication

It is necessary for employers and organizational IT administrators to provide background knowledge and risk trade-offs of non-adoption before deploying MFA. In addition, the majority of users do not properly understand the risk of single-factor authentication nor do they gain sufficient security awareness, as they stated multi-factor is "unnecessary" and "a waste of time" for them. Hence, necessary risk communication is required for end users. For instance, lots of users showed a lack of understanding of authentication factors. Defining authentication factors for users, such as "what you have" and "what you know," will let users understand the attack surface of each factor and the security benefits of MFA. In addition, we observed users' concerns with setting up the application for use. This is likely to happen with MFA solutions that target enterprise markets, as many personal services streamline design to improve the MFA on-boarding experience. It is a good idea for system administrators to provide detailed step-by-step instructions for setting up MFA, which shows significant adoption improved in pas research (Das, Russo, Dingman, Dev, Kenny & Camp 2018).

## 5.2. Backup and Migration Improvement

To eliminate the concerns for disaster recovery, properly designed secure and convenient backup and migration systems should be implemented in MFA solutions. While it is necessary to maintain device ownership as an important authentication factor, technologies, such as near-field communication (NFC) between two devices, can streamline the procedure for migration: in such a case, the original device authenticates the ownership of the new device, then proceeds to generate and store new seeds for MFA on the new device. More recovery mechanisms for lost devices should also be implemented, aside from recovery passwords. Although routine checks for recovery mechanisms is necessary to prevent loss of control over accounts, it should be designed in a non-interrupting way, since the main motivation of a user opening authentication apps is to retrieve authentication credentials quickly. It is feasible to send background notifications regularly to let users verify the status of recovery mechanisms.

### 5.3. Application Development and Testing

More application and system integration testing is required to ensure that the MFA experience is properly delivered. System administrators should conduct “pilot testing" with feedback mechanisms before rolling out MFA to the organization at large. Application and service vendors should keep track of device updates so that they can deliver necessary device support to customers.

## 6. Conclusion

Multi-factor authentication is an important initiative for information security in modern online applications. It mitigates the risk of password breaches in an age of frequent online attacks. However, current MFA implementations have not yet achieved the state of general usability, and users are unwilling to enroll in MFA unless required to do so by organizational policies. Through our detailed user review analysis and recommendations, we aim to provide security for all. We conclude that, MFA technologies should be designed in a more elegant and effortless way to relieve users’ concerns regarding device dependency, applications’ ease of use, backup and migration issues, and provide proper risk communication and user training.

## 7. Acknowledgment

This research was supported in part by the National Science Foundation under CNS 1565375, Cisco Research Support, and the Comcast Innovation Fund. Any opinions, findings, and conclusions or recommendations expressed in this material are those of the author(s) and do not necessarily reflect the views of the US Government, the National Science Foundation, Cisco, Comcast, nor Indiana University. We would like to thank Andrew Kim, Joshua Streiff, and Olivia Kenny for providing feedback in completion of this paper.

## References

Althobaiti, M. M. & Mayhew, P. (2014), Security and usability of authenticating process of online banking: User experience study, *in* ‘Security Technology (ICCST), 2014 International Carnahan Conference on’, IEEE, pp. 1–6.

Amin, A., ul Haq, I. & Nazir, M. (2017), ‘Two factor authentication’, *International Journal of Computer Science and Mobile Computing.*

Armington, J. & Ho, P. (2003), ‘Robust multi-factor authentication for secure application environments’. US Patent App. 10/086,123.

Beck, K., Beedle, M., Van Bennekum, A., Cockburn, A., Cunningham, W., Fowler, M., Grenning, J., Highsmith, J., Hunt, A., Jeffries, R. et al. (2001), ‘Manifesto for agile software development’.

Braz, C. & Robert, J.-M. (2006), Security and usability: the case of the user authentication methods, *in* ‘IHM’, Vol. 6, pp. 199–203.

Camp, L. J., Abbott, J. & Chen, S. (2016), Cpasswords: Leveraging episodic memory and human-centered design for better authentication, *in* '2016 49th Hawaii International Conference on System Sciences (HICSS)', IEEE, pp. 3656–3665.

Choong, Y.-Y., Theofanos, M. & Liu, H.-K. (2014), *United States Federal Employees' Password Management Behaviors: A Department of Commerce Case Study*, US Department of Commerce, National Institute of Standards and Technology.

Colnago, J., Devlin, S., Oates, M., Swoopes, C., Bauer, L., Cranor, L. & Christin, N. (2018), "it's not actually that horrible": Exploring adoption of two-factor authentication at a university, *in* 'Proceedings of the 2018 CHI Conference on Human Factors in Computing Systems', ACM, p. 456.

Das, S., Dingman, A. & Camp, L. J. (2018), Why johnny doesn't use two factor a twophase usability study of the fido u2f security key, *in* '2018 International Conference on Financial Cryptography and Data Security (FC)'.

Das, S., Russo, G., Dingman, A. C., Dev, J., Kenny, O. & Camp, L. J. (2018), A qualitative study on usability and acceptability of yubico security key, *in* 'Proceedings of the 7th Workshop on Socio-Technical Aspects in Security and Trust', ACM, pp. 28–39.

Das, S., Kim, A., Tingle, Z. & Nipprt-Eng, C. (2019), All about phishing exploring user research through a systematic literature review, *in* 'Proceedings of the Thirteenth International Symposium on Human Aspects of Information Security & Assurance (HAISA 2019)'.

Das, S., Wang, B., Tingle, Z. & Camp, L.J (2019), Evaluating User Perception of Multi-Factor Authentication A Systematic Review, *in* 'Proceedings of the Thirteenth International Symposium on Human Aspects of Information Security & Assurance (HAISA 2019)'.

Ding, X., Liu, B. & Yu, P. S. (2008), A holistic lexicon-based approach to opinion mining, *in* 'Proceedings of the 2008 international conference on web search and data mining', ACM, pp. 231–240.

Fu, B., Lin, J., Li, L., Faloutsos, C., Hong, J. & Sadeh, N. (2013), Why people hate your app: Making sense of user feedback in a mobile app store, *in* 'Proceedings of the 19th ACM SIGKDD International Conference on Knowledge Discovery and Data Mining', KDD '13, ACM, New York, NY, USA, pp. 1276–1284.

Furnell, S. (2007), 'A comparison of website user authentication mechanisms', *Computer Fraud & Security* 2007(9), 5–9.

Hwang, M.-S. & Li, L.-H. (2000), 'A new remote user authentication scheme using smart cards', *IEEE Transactions on consumer Electronics* 46(1), 28–30.

Jindal, N. & Liu, B. (2008), Opinion spam and analysis, *in* 'Proceedings of the 2008 international conference on web search and data mining', ACM, pp. 219–230.

Joyce, R. (2016), 'Disrupting nation state hackers', *USENIX Enigma. San Fransisco, CA.*

Li, F. H., Huang, M., Yang, Y. & Zhu, X. (2011), Learning to identify review spam, *in* 'Twenty-Second International Joint Conference on Artificial Intelligence'.

McCoy, D., Bauer, K., Grunwald, D., Tabriz, P. & Sicker, D. (2007), 'Shining light in dark places: A study of anonymous network usage', *University of Colorado Technical Report CU-CS-1032-07 (August 2007)* .

Md Noman, A. S., Das, S. & Patil, S. (2019), Rejected by techies: Understanding facebook non-adoption by experts via user generated content, *in* 'Proceedings of the 2019 CHI Conference on Human Factors in Computing Systems', ACM.

Mukherjee, A., Liu, B. & Glance, N. (2012), Spotting fake reviewer groups in consumer reviews, *in* 'Proceedings of the 21st international conference on World Wide Web', ACM, pp. 191–200.

Pang, B., Lee, L. et al. (2008), 'Opinion mining and sentiment analysis', *Foundations and Trends* R *in Information Retrieval* 2(1–2), 1–135.

Ting, D. M., Hussain, O. & LaRoche, G. (2015), 'Systems and methods for multi-factor authentication'. US Patent 9,118,656.

Vasa, R., Hoon, L., Mouzakis, K. & Noguchi, A. (2012), A preliminary analysis of mobile app user reviews, *in* 'Proceedings of the 24th Australian Computer-Human Interaction Conference', ACM, pp. 241–244.

Ward, M. A. (2006), 'Information systems technologies: A public-private sector comparison', *Journal of Computer Information Systems* 46(3), 50–56.

Weir, C. S., Douglas, G., Richardson, T. & Jack, M. (2010), 'Usable security: User preferences for authentication methods in ebanking and the effects of experience', *Interacting with Computers* 22(3), 153–164.

Whitten, A. & Tygar, J. D. (1999), Why johnny can't encrypt: A usability evaluation of pgp 5.0., *in* 'USENIX Security Symposium', Vol. 348.